\documentclass[preprint,aps,amsmath,amssymb]{revtex4} 

\usepackage{graphicx}
\usepackage{dcolumn}
\usepackage{bm}

\newcommand{\mymass}{$2697.5\,\pm\,2.2$~MeV$/c^2$} 
\newcommand{\mylife}{$72\,\pm\,11\,\pm\,11$~fs} 
\newcommand{\myield}{$64\,\pm\,14$}

\begin{document}

\title{Measurement of the $\Omega_c^0$ Lifetime}

\date{\today}

\affiliation{University of California, Davis, CA 95616}
\affiliation{Centro Brasileiro de Pesquisas F\'isicas, Rio de Janeiro, RJ, Brasil}
\affiliation{CINVESTAV, 07000 M\'exico City, DF, Mexico}
\affiliation{University of Colorado, Boulder, CO 80309}
\affiliation{Fermi National Accelerator Laboratory, Batavia, IL 60510}
\affiliation{Laboratori Nazionali di Frascati dell'INFN, Frascati, Italy I-00044}
\affiliation{University of Illinois, Urbana-Champaign, IL 61801}
\affiliation{Indiana University, Bloomington, IN 47405}
\affiliation{Korea University, Seoul, Korea 136-701}
\affiliation{Kyungpook National University, Taegu, Korea 702-701}
\affiliation{INFN and University of Milano, Milano, Italy}
\affiliation{University of North Carolina, Asheville, NC 28804}
\affiliation{Dipartimento di Fisica Nucleare e Teorica and INFN, Pavia, Italy}
\affiliation{University of Puerto Rico, Mayaguez, PR 00681}
\affiliation{University of South Carolina, Columbia, SC 29208}
\affiliation{University of Tennessee, Knoxville, TN 37996}
\affiliation{Vanderbilt University, Nashville, TN 37235}
\affiliation{University of Wisconsin, Madison, WI 53706}
\author{J.~M.~Link}
\affiliation{University of California, Davis, CA 95616}
\author{M.~Reyes}
\affiliation{University of California, Davis, CA 95616}
\author{P.~M.~Yager}
\affiliation{University of California, Davis, CA 95616}
\author{J.~C.~Anjos}
\affiliation{Centro Brasileiro de Pesquisas F\'isicas, Rio de Janeiro, RJ, Brasil}
\author{I.~Bediaga}
\affiliation{Centro Brasileiro de Pesquisas F\'isicas, Rio de Janeiro, RJ, Brasil}
\author{C.~G\"obel}
\affiliation{Centro Brasileiro de Pesquisas F\'isicas, Rio de Janeiro, RJ, Brasil}
\author{J.~Magnin}
\affiliation{Centro Brasileiro de Pesquisas F\'isicas, Rio de Janeiro, RJ, Brasil}
\author{A.~Massafferri}
\affiliation{Centro Brasileiro de Pesquisas F\'isicas, Rio de Janeiro, RJ, Brasil}
\author{J.~M.~de~Miranda}
\affiliation{Centro Brasileiro de Pesquisas F\'isicas, Rio de Janeiro, RJ, Brasil}
\author{I.~M.~Pepe}
\affiliation{Centro Brasileiro de Pesquisas F\'isicas, Rio de Janeiro, RJ, Brasil}
% ...
\collaboration{The FOCUS Collaboration}
\affiliation{See \textrm{http://www-focus.fnal.gov/authors.html} for additional author information.}

\begin{abstract}

The FOCUS experiment(FNAL-E831) has used two channels, $\Omega^- \pi^+$ and
$\Xi^-K^- \pi^+ \pi^+$,
to measure the lifetime of the $\Omega_c^0$ charmed baryon. 
From a sample of $64 \pm 14$ signal events at a mass of 2.698~GeV/$c^2$, we measure an $\Omega_c^0$ 
lifetime of $72 \pm 11$\,(stat.)~$\pm\:11$\,(sys.)~fs, substantially improving upon the current world
average.
\end{abstract}

\maketitle

\section{\label{sec:level1}Introduction}

Several experiments have searched for the $\Omega_c^0$, $J^P = 1/2^+ \{css\}$ ground state. 
The first claim of an observation of the $\Omega_c^0$ was made by  CERN experiment
WA62 with a cluster of 3 events in the decay channel of $\Omega_c^0 \rightarrow \Xi^- K^- \pi^+ \pi^+$ 
(throughout this Letter, charge conjugate states are assumed)
at a mass of $2740 \pm 20$~MeV$/c^2$ \cite{Z.phys.C28.175}.
The ARGUS collaboration followed with signals for the  $\Omega_c^0$ in $\Omega^- \pi^+ \pi^+ \pi^-$ and 
$\Xi^- K^- \pi^+ \pi^+$ based on $0.380$~fb$^{-1}$ of data \cite{APS.1993}, 
but these signals were not confirmed by  the CLEO experiment which had a much higher sensitivity. 
Fermilab photoproduction experiment E687 reported an observation of the $\Omega_c^0$ decaying to $\Sigma^+ K^- K^- \pi^+$
with a mass of $2699.9 \pm 1.5 \pm 2.5$~MeV$/c^2$ \cite{PLB.338.106} and a lifetime of 
$86~^{+27}_{-20} \pm 28$~fs~\cite{PLB.357.678}. 
E687 published an earlier observation in the $\Omega^- \pi^+$ channel with a mass 
of $2705.9 \pm 3.3 \pm 2.3$~MeV$/c^2$~\cite{PLB.300.190}. 
In 1995, CERN experiment WA89 reported 200 $\Omega_c^0$ events in seven modes, although the published 
lifetime result of $55~^{+13}_{-11}~^{+18}_{-23}$~fs comes from only two of the decay modes~\cite{PLB.358.151}.
In 2000, CLEO presented an $\Omega_c^0$ mass of $2694.6 \pm 2.6 \pm 1.9$ MeV$/c^2$ with the combined signal from 
four decay modes~\cite{PRL.86.3730}. 

Clearly, the lifetime measurement of the $\Omega_c^0$ is still not well measured.
Additional measurements with improved statistical accuracy are needed to test 
theoretical models.
The lifetime measurement is particularly important (when combined with other charm baryon lifetime
measurements) in estimating the interference effects from different contributing 
diagrams \cite{PLB.289.455}.  The current uncertainty on the 
$\Omega_c^0$ lifetime is more than 30\%~\cite{Phys.Rev.D66} of the lifetime value and is too large to extract 
meaningful information on the interfering amplitudes. In this Letter we report
a new lifetime value of the $\Omega_c^0$ baryon from the FOCUS experiment.

The FOCUS spectrometer is well-suited to reconstruct short-lived charm decays.
Two silicon microvertex systems provide excellent separation between the production
and charm decay vertices.  The target silicon system (TS) consists of two pairs of silicon planes,
each immediately downstream of a pair of BeO target segments.  The second silicon strip detector (SSD)
consists of 12 silicon planes, downstream of the target region.
Charged particles are tracked and momentum analyzed with five
stations of multiwire proportional chambers in a two magnet forward spectrometer. Three
multicell threshold \v{C}erenkov detectors are used to identify electrons, pions, kaons,
and protons and are described in detail in a previous FOCUS publication~\cite{hep.ex.0108011}.

\section{\label{sec:level2}Reconstruction of hyperons, $\Xi^-$ and $\Omega^-$ }

A detailed description of $\Xi^-$ and $\Omega^-$ reconstruction in the FOCUS
spectrometer can be found elsewhere~\cite{hep.ex.0109028}. 
Using the ``cascade'' reconstruction algorithm, 
we are able to reconstruct the decays $\Xi^- \!\rightarrow\!  \Lambda^0 \pi^-$ and 
$\Omega^- \!\rightarrow\! \Lambda^0 K^-$, 
which have branching fractions of 99.9\% and 67.8\%, respectively. 
In this analysis, we only use $\Xi^-$'s and $\Omega^-$'s which decay downstream of the SSD.
This allows us to track the $\Xi^-$ or $\Omega^-$ in the SSD before it decays. 
A vertex is found between a $\Lambda^0$ and a $\pi^-$ or a $K^-$ and the
 $\Lambda^0 \pi^-$ or $\Lambda^0 K^-$ combined
momentum vector must match the slopes and positions of a track in the SSD.
For $\Omega^-$ candidates, we require mass differences of
$|M(\Xi^-) - M(\Lambda^0 \pi^-)| > 30$~MeV$/c^2$ and 
$|M(\Omega^-) - M(\Lambda^0 K^-)| < 20$~MeV$/c^2$, ensuring that  
most of the more copiously produced $\Xi^-$'s are removed from the $\Omega^-$ sample. 

\section{\label{sec:level3}Reconstruction of $\Omega_c^0$ candidates} 
\begin{figure}[ht]
  \begin{center}
     \includegraphics[width=7in]{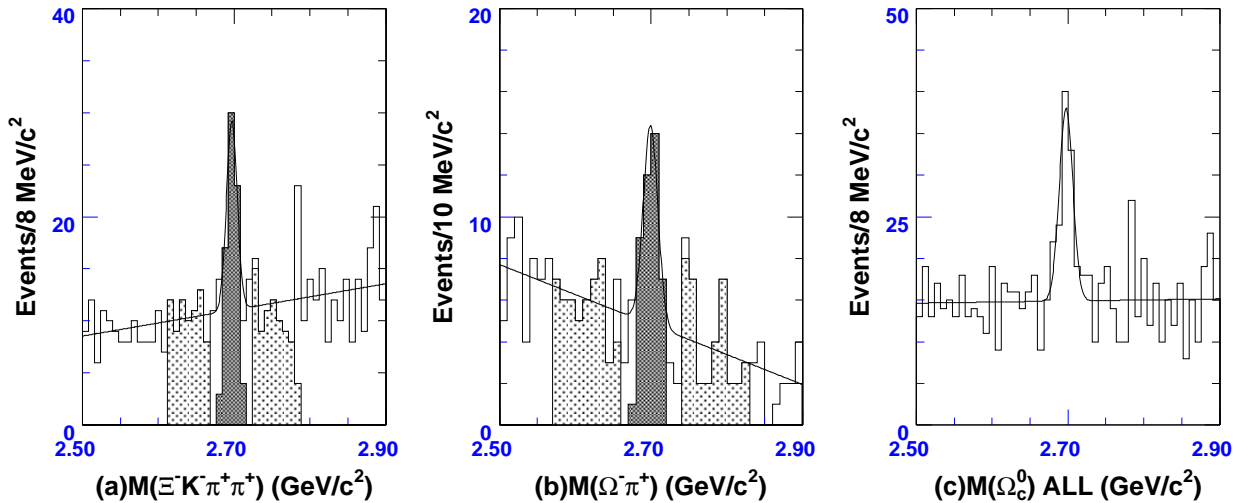}
  \caption{Invariant mass distributions for $\Omega_c^0$ candidates: 
    (a) Reconstructed mass of $\Xi^- K^- \pi^+ \pi^+$. There are $38 \pm 9$ events 
at a mass of $2696.5 \pm 1.9$~MeV$/c^2$.  
    (b) Reconstructed mass of $\Omega^- \pi^+ $. There are $23 \pm 7$ events at 
a mass of $2699.4 \pm 3.4$~MeV$/c^2$.  
    (c) Combined invariant mass distribution. There are \myield\ events at a mass 
of \mymass.
We define the signal region (hatched area) to be
        within $2\,\sigma$ of the fitted mass value and the two sideband regions 
(dotted area) are 4--12\,$\sigma$ from the fitted mass value.
   }
  \label{fig:mass}
\end{center}
\end{figure}

The $\Omega_c^0$ candidates are formed by making a vertex hypothesis for the daughter 
particles.
We use a $\Xi^-$ and three charged tracks of the right charge combination 
for the $\Xi^- K^- \pi^+ \pi^+$ mode, and  
 an $\Omega^-$ and an oppositely charged track for the $\Omega^-\pi^+$ mode.
The confidence level of the decay vertex of the $\Omega_c^0$ candidate is 
required to be greater than 10\%.  The combined momentum vector located at the 
decay vertex forms the $\Omega_c^0$ track.
A candidate driven vertexing algorithm~\cite{nim.320.519} uses the 
$\Omega_c^0$ track as a seed track to find a production vertex with a
confidence level greater than 1\%.  The primary multiplicity, including the 
seed track, must be at least 3 tracks and the production vertex must be 
inside a target.
The significance of separation between the production and the decay vertices 
($L/\sigma_L$) must be greater than 2 
for the $\Xi^- K^- \pi^+ \pi^+$ mode and greater than 0 for the $\Omega^- \pi^+$ mode. 
Different values are chosen for the $L/\sigma_L$ cut due to a difference in the secondary 
vertex resolution.
\v{C}erenkov particle identification (PID) is performed by constructing a 
log likelihood value ${\mathcal W}_i$ for the particle 
hypotheses ($i = e, \pi, K, p$ ). 
The $\pi$ consistency of a track is defined by $\Delta {\mathcal W}_\pi = {\mathcal W}_\textrm{min} - {\mathcal W}_\pi$, 
where ${\mathcal W}_\textrm{min}$ is the minimum ${\mathcal W}$ value of 
the other three hypotheses. Similarly, we define $\Delta {\mathcal W}_{K,\pi} = {\mathcal W}_\pi - {\mathcal W}_K$ 
for kaon identification. We 
require $\Delta {\mathcal W}_{K,\pi} > 3 $ for kaons 
and $\Delta {\mathcal W}_\pi > -6$ for pions.
In the $\Xi^- K^- \pi^+ \pi^+$ mode, we add additional combination PID cuts based on
Monte Carlo simulation studies. We define $\sum \Delta {\mathcal W}_\pi$ 
to be the positive sum over pion candidates and the 
negative sum over other particle candidates. Also $\Delta {\mathcal W}_{K,\pi} + \Delta {\mathcal W}_{K,p}$ is 
used  for separation between the protons and the kaons.     
$\sum \Delta {\mathcal W}_\pi$ and $\Delta {\mathcal W}_{K,\pi} + \Delta {\mathcal W}_{K,p}$ 
are required to be greater than 7. 
In the $\Omega^- \pi^+ $ mode the momentum asymmetry,  
$(P_{\Omega^-} - P_{\pi^+})/(P_{\Omega^-} + P_{\pi^+})$, is required to be  
greater than $-0.2$ and less than $0.7$.  
Also the $\pi^+$ transverse momentum must be larger than $0.2~$~GeV$/c$ and the
the momentum of $\Omega_c^0$ must be greater than $50~$~GeV$/c$.
The resulting mass spectra are shown in  Fig.~\ref{fig:mass}.  The mass 
spectra are fit with a Gaussian function for the signal distribution and 
a first order polynomial function for the background.  From the combined sample, we
find a fitted mass of \mymass\ (systematic uncertainty not evaluated), consistent with the results of
other experiments.

\section{\label{sec:level4}Lifetime measurement} 

To measure a lifetime in fixed target experiments, we use a binned maximum 
likelihood technique~\cite{PLB.268.584}.  We fit the reduced proper time 
distribution, defined as $t'=(L-N\sigma_L)/\beta\gamma c = t-N\sigma_t$,
where $N$ is the separation cut value between the production and the decay vertex,
$\beta c$ is the particle velocity, and $\gamma$ is the Lorentz boost factor to 
the $\Omega_c^0$ center of mass frame. 

\begin{figure}[ht]
  \begin{center}
     \includegraphics[width=6.5in]{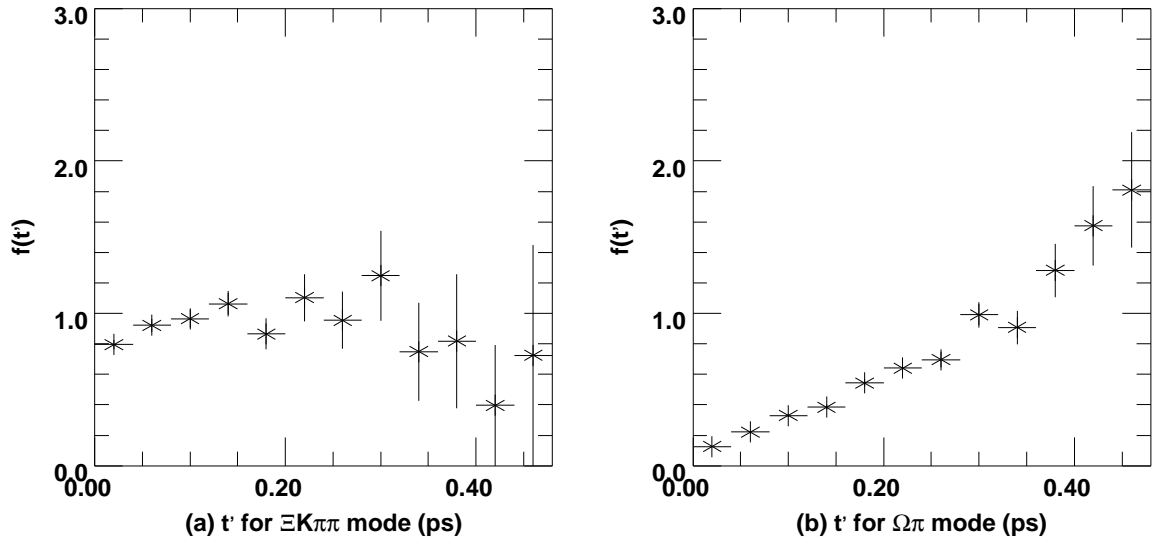}
  \caption{The $f(t^{\prime})$ correction function is displayed for each mode.}
  \label{fig:ft}
\end{center}
\end{figure}

The signal region is defined to lie within $2\,\sigma$ of the fitted $\Omega_c^0$ mass. 
The background is assumed to have the same lifetime behavior in the signal region as
in the sidebands, 4--12\,$\sigma$ away from the peak.
Taking $S$ as the number of signal events in the signal region and $B$ as the total
number of background events in the same region, the expected number of events $n_i$ in
the $i^{th}$ reduced proper time bin centered at $t'_i$ is given by:
\begin{equation}
n_{i} = S\frac{f(t'_{i})e^{-t'_{i}/\tau}}
                 { \displaystyle \sum_{ i}^{} f(t'_{i})e^{-t'_{i}/\tau}  }+
           B\frac{b_{i}}{\displaystyle \sum_{i}^{} b_{i}}
\label{eq:ni}
\end{equation}
where $b_i$ describes the background reduced proper time as
estimated from sidebands and $f(t'_{i})$ is a correction function
which takes into account the effects of spectrometer acceptance and efficiency, 
analysis cut efficiencies, and particle absorption.  The $f(t')$ distribution are 
shown in Fig.~\ref{fig:ft} for each decay mode.

\begin{figure}[ht]
  \begin{center}
     \includegraphics[width=6.5in]{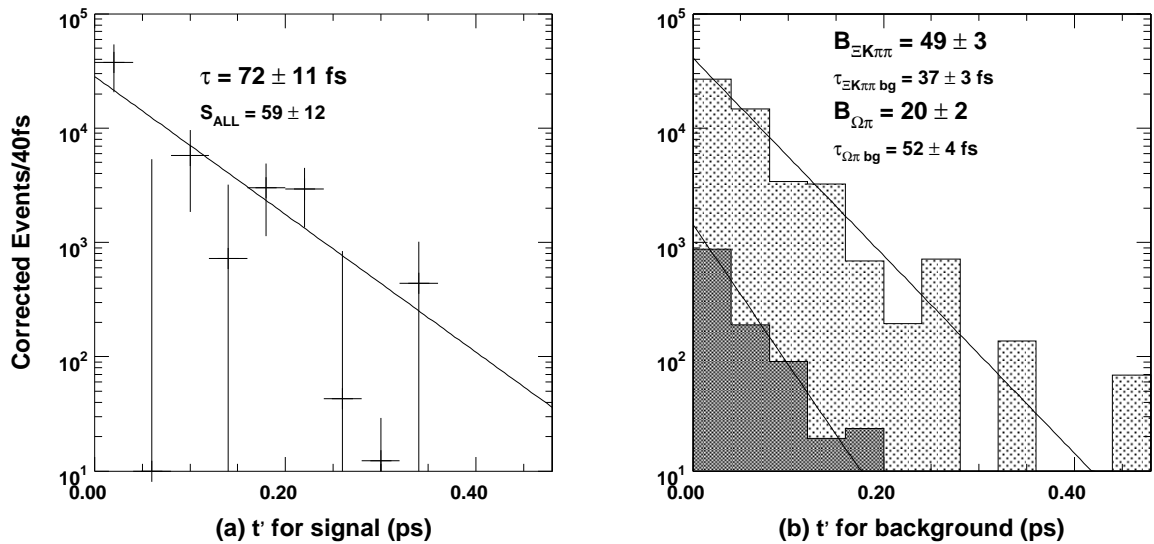}
  \caption{(a) The corrected $t'$ distribution with the lifetime fit function for the combined signal.
(b) The $t'$ distributions of expected backgrounds in the signal band for each mode; 
the dark region is for $\Xi^- K^- \pi^+ \pi^+$ and the light one is for $\Omega^- \pi^+$.  
Lines show the lifetime fitting functions for signal and background 
distributions. 
The lifetime fit finds $59 \pm 12$ signal events rather than $64 \pm 14$ due to the $ 2 \sigma$ mass window used.}
  \label{fig:lifetime}
\end{center}
\end{figure}

\begin{figure}[ht]
  \begin{center}
     \includegraphics[width=6.5in]{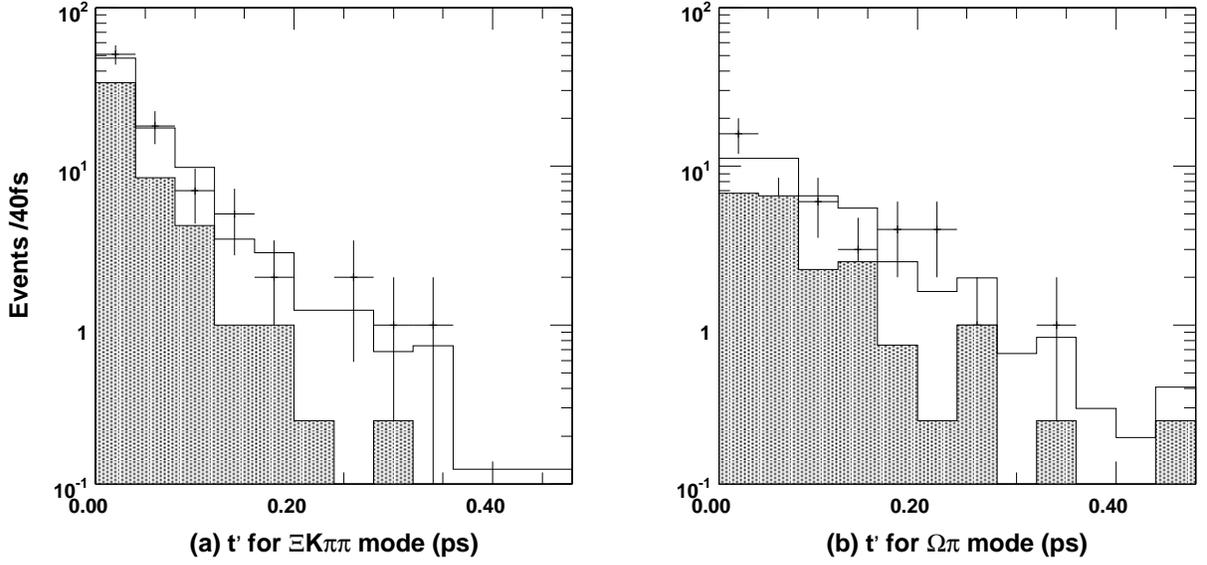}
  \caption{ The predicted events (histogram) are superimposed on the observed events (points)
while the shaded distribution displays the $t'$ distribution of the
background for (a) the ${\Xi^- K^- \pi^+ \pi^+}$ mode and (b) the ${\Omega^- \pi^+}$ mode.}
  \label{fig:mc}
\end{center}
\end{figure}

The likelihood is constructed from the product of the Poisson probability of observing
$s_i$ events when $n_i$ are expected with the Poisson
probability of observing $N_b = \sum b_{i}$ events in the sidebands when 4$B$ background
events 
are expected. The factor of 4 accounts for the fact that the sideband region is
four times wider than the signal region.
The likelihood takes the form:
\begin{equation}
{\mathcal L}= \left( \prod_{i} \frac{n_{i}^{s_{i}}e^{-n_{i}}}{s_{i}!}\right)
  \times
  \left(\frac{(4B)^{\displaystyle N_b}e^{-4B}}
  {\displaystyle N_b!}\right)
\end{equation}
The combined likelihood function is given by the product of the likelihoods:
\begin{equation}\begin{split}
 {\mathcal L}_{\Omega_c^0}=
    {\mathcal L}_{\Xi^- K^- \pi^+ \pi^+}      \times
    {\mathcal L}_{\Omega^- \pi^+}             
 \end{split}\label{eq:hood}
 \end{equation}
There are 3 fit parameters; one parameter for the lifetime $\tau$ and
two parameters,  B$_{\Xi^- K^- \pi^+ \pi^+}$ and B$_{\Omega^- \pi^+}$, for the backgrounds 
from each mode.
Our measurement of the $\Omega_c^0$ lifetime is $72 \pm 11$~fs as shown in Fig. \ref{fig:lifetime}. 
In Fig. \ref{fig:mc}, the $t'$ distributions from the data and from the fit are compared with each other.  

\section{\label{sec:level5}Studies of systematic errors} 

\begin{figure}[ht]
  \begin{center}
     \includegraphics[width=6.5in]{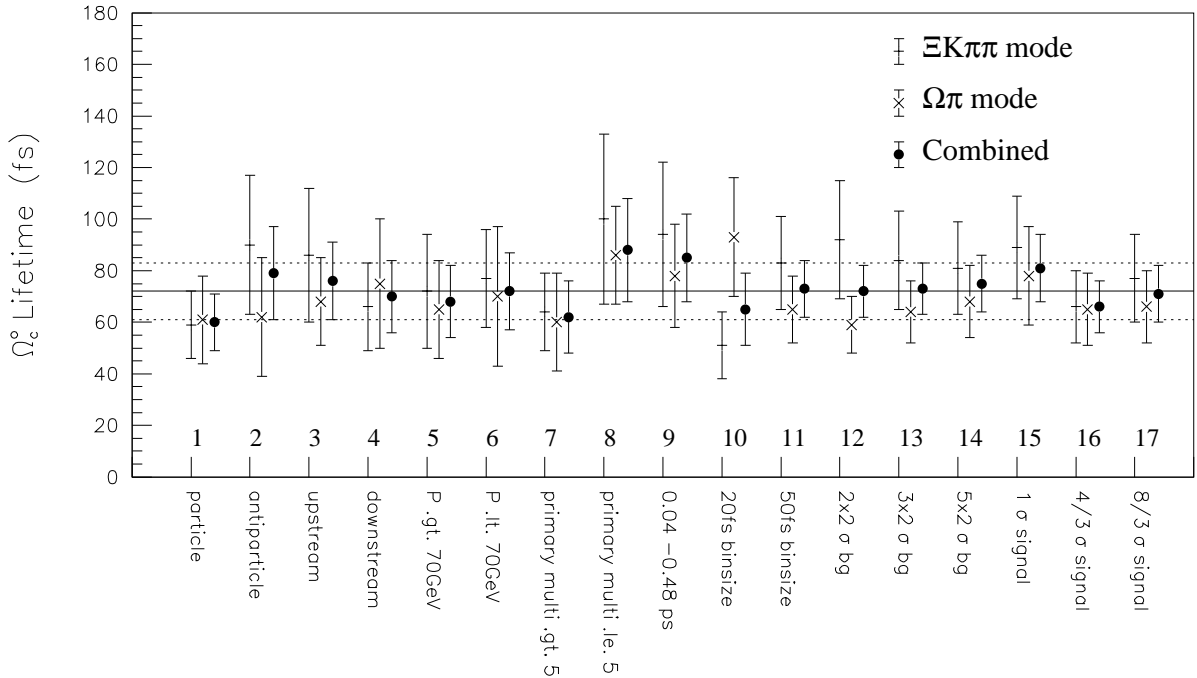}
  \caption{ Lifetime measurements for systematic studies. 
The solid line represents the best determined value for the $\Omega^0_c$ lifetime 
and the two dotted lines show the extent of the statistical error.  
   }
  \label{fig:sys-error}
\end{center}
\end{figure}

\begin{table}
\begin{center}
\caption{ The itemized list of the systematic uncertainties. The numbers in the Items column refer to 
the entry numbers in Fig.~\ref{fig:sys-error}.} 
\begin{tabular}{ccc}
\hline\hline
\rule[-0.3cm]{0pt}{0.8cm}
Systematic Source  & Items & Uncertainty (fs) \\
\hline
\rule[-0.3cm]{0pt}{0.5cm}
Split Sample Method    & Particle and antiparticle (1,2)      & $\sim 0$  \\ 
                       & Upstream and downstream target (3,4) & $\sim 0$  \\  
                       & High and low momentum (5,6) & $\sim 0$  \\
                       & Primary vertex multiplicity (7,8) & $\sim 0$  \\ 
Fit variant            & Bin size (10,11) and fitting region (9)  & $ \pm 9$  \\ 
$t'$ Resolution        & Toy Monte Carlo studies              & $ \pm 4$  \\ 
Background             & Sideband (12,13,14) and signal band (15,16,17)& $ \pm 5$  \\ 
\hline
Total                  &                                      & $ \pm 11$  \\ \hline \hline
\end{tabular}
\label{tb:sys-error}
\end{center}
\end{table}

\begin{table}
\begin{center}
\caption{ The $\Omega^0_c$ lifetime measurements}
\begin{tabular}{ccc}
\hline\hline
\rule[-0.3cm]{0pt}{0.8cm}
Experiment  & Lifetime & Decay Modes \\
\hline
\rule[-0.3cm]{0pt}{0.5cm}
E687                   & 86 $^{+27}_{-20} \pm$ 28 fs       & $\Sigma^+K^-K^-\pi^+$  \\
WA89                   & 55 $^{+13}_{-11}$ $^{+18}_{-23}$ fs & $\Xi^-K^-\pi^+\pi^+$, $\Omega^-\pi^+\pi^+\pi^-$ \\
PDG2002                & 64 $\pm$ 20 fs                    &   \\
FOCUS                  & 72 $\pm$ 11 $\pm$ 11 fs            & $\Xi^-K^-\pi^+\pi^+$, $\Omega^-\pi^+$  \\
\hline \hline
\end{tabular}
\label{tb:compare}
\end{center}
\end{table}

We have studied various systematic uncertainties associated with the Monte Carlo modeling 
by computing the lifetimes of 
independent data samples split by particle/antiparticle, primary vertex position
(upstream and downstream target region), $\Omega_c^0$ momentum 
(greater than 70 GeV$\!/\!\!\:c$ / less than 70 GeV$\!/\!\!\:c$) and production vertex multiplicity ($>$5 / $\le 5$). 
All lifetimes from these samples are consistent within the statistical error as 
shown in Fig.~\ref{fig:sys-error} points 1--8, indicating a negligible systematic error due to the
Monte Carlo simulation.

Since the binned likelihood method has been used to measure the lifetime, we have 
investigated the uncertainty from the fit range and binning effects by examining the 
variance in lifetime for different $t'$ bin sizes (Fig.~\ref{fig:sys-error} points 10--11) and 
for different fitting ranges (Fig.~\ref{fig:sys-error} point 9). 

The proper time resolution of our fully simulated Monte Carlo for the $\Omega_c^0$ data is about 
40--50 fs. We have tested the accuracy of the fitting procedure when the
lifetime is comparable to the proper time resolution. We used a toy Monte Carlo study to test
the fitting procedure using the proper time from
which we extracted a systematic uncertainty of 4~fs. The toy Monte Carlo test was also
used to validate the statistical error determination.

The systematic uncertainty due to the background contamination is examined by investigating the reflections 
from other charm baryon decays and by varying the sideband
and signal regions (Fig.~\ref{fig:sys-error} points 12--17). 

The studies of the systematic uncertainties are summarized in Table~\ref{tb:sys-error}.  
The total systematic uncertainty of the $\Omega_c^0$ lifetime measurement is determined to 
be 11~fs by adding all of the systematic uncertainties in quadrature.    

\section{\label{sec:level6}Conclusion}
We measure an $\Omega_c^0$ lifetime of \mylife\ using \myield\ events in the 
two decay modes, $\Omega^- \pi^+$ and
$\Xi^-K^- \pi^+ \pi^+$. 
We compare our result with previous measurements 
in Table~\ref{tb:compare}. Our lifetime result is consistent with
previous $\Omega_c^0$ lifetime results.  This 20\% measurement of the lifetime
substantially improves upon the current (30\%) world average.

\section{\label{sec:level8}Acknowledgements} 
~
We wish to acknowledge the assistance of the staffs of Fermi National
Accelerator Laboratory, the INFN of Italy, and the physics departments
of
the
collaborating institutions. This research was supported in part by the
U.~S.
National Science Foundation, the U.~S. Department of Energy, the Italian
Istituto Nazionale di Fisica Nucleare and 
Ministero della Istruzione, Universit\`a e
Ricerca, the Brazilian Conselho Nacional de
Desenvolvimento Cient\'{\i}fico e Tecnol\'ogico, CONACyT-M\'exico, and
the Korea Research Foundation of the  
Korean Ministry of Education.

\end{document}